\documentclass{emulateapj}
\usepackage{natbib}
\usepackage{lscape} 

\newcommand{\eg}{e.g., }

\newcommand{\Msun}{M_{\odot}}

\newcommand{\Cofs}{$^{56}$Co}
\newcommand{\Nifs}{$^{56}$Ni}

\newcommand{\Mej}{M_{\rm ej}}
\newcommand{\KE}{E_{\rm K}}

\def\gsim{\mathrel{\rlap{\lower 4pt \hbox{\hskip 1pt $\sim$}}\raise 1pt
\hbox {$>$}}}
\def\lsim{\mathrel{\rlap{\lower 4pt \hbox{\hskip 1pt $\sim$}}\raise 1pt
\hbox {$<$}}}

\shorttitle{Broad line Type Ic SN 2007ru}
\shortauthors{Sahu et al.}
 
\begin{document}

\title{The broad line Type Ic supernova SN 2007ru: Adding to the diversity of
Type Ic supernovae }
\author{
D.K. Sahu\altaffilmark{1},
Masaomi Tanaka\altaffilmark{2},
G.C. Anupama\altaffilmark{1},
Uday K. Gurugubelli\altaffilmark{1,3}, and 
Ken'ichi Nomoto\altaffilmark{4,2} }

\altaffiltext{1}{Indian Institute of Astrophysics, II Block Koramangala,
Bangalore 560034, India: dks@iiap.res.in, gca@iiap.res.in, uday@iiap.res.in}
\altaffiltext{2}{Department of Astronomy, Graduate School of Science,
University of Tokyo, Hongo 7-3-1, Bunkyo-ku, Tokyo 113-0033, Japan: mtanaka@astron.s.u-tokyo.ac.jp }
\altaffiltext{3}{Joint Astronomy Programme, Indian Institute of Science, Bangalore 560012, India}
\altaffiltext{4}{Institute for the Physics and Mathematics of the
Universe, University of Tokyo, Kashiwa, Chiba 277-8582, Japan: nomoto@astron.s.u-tokyo.ac.jp}

\begin{abstract}
Photometric and spectral evolution of the Type Ic supernova SN 2007ru 
until around 210 days after maximum are presented. 
The spectra show broad spectral 
features due to very high expansion velocity, normally seen in hypernovae. 
The photospheric velocity is higher than other normal Type Ic supernovae. 
It is lower than SN 1998bw at $\sim$ 8 days after the explosion, but is comparable 
at later epochs. The light curve evolution 
of SN 2007ru indicates a fast rise time of 8$\pm$3 days to $B$ band maximum 
and post-maximum decline more rapid  than other broad-line Type Ic supernovae.
With an absolute $V$ magnitude of $-19.06$, SN 2007ru is comparable in 
brightness with SN 1998bw and lies at the  brighter end of the observed Type 
Ic supernovae. The ejected mass of \Nifs\ is estimated to be  $\sim 0.4\Msun$. The 
fast rise and decline of the light curve and the high expansion velocity
suggest that SN 2007ru is an explosion with a high kinetic energy/ejecta mass
ratio ($E_{\rm K}/M_{\rm {ej}}$). This adds to the diversity of Type Ic supernovae. 
Although the early phase spectra are most similar to 
those of broad-line SN 2003jd, the [OI] line profile in the nebular spectrum of SN 2007ru
shows the singly-peaked profile, in contrast to the doubly-peaked profile in SN 2003jd.
The singly-peaked profile, together with 
the high luminosity and the high expansion velocity, may suggest that
SN 2007ru could be an aspherical explosion viewed from the polar direction.
Estimated  oxygen  abundance  12 + log(O/H) of $\sim$8.8 indicates that SN 2007ru 
occurred in
a region with nearly solar metallicity.
\end{abstract}

\keywords{supernovae: general supernovae: individual (SN 2007ru) }


\section{Introduction}
Broad-line Type Ic supernovae (SNe Ic) are a subclass of core collapse
SNe Ic that have broad features in their spectra,
indicating unusually high expansion velocities reaching close to 0.1$c$ 
at early times. Only a few candidates 
of this class are known. Some broad-line SNe Ic are associated 
with Gamma Ray Bursts (GRBs)
(Galama et al. 1998; Matheson et al. 2003; Malesani et al. 2004), 
or X-Ray Flash (XRF) (Pian et al. 2006; Modjaz et al. 2006), 
while some others do not show any clear evidence of being associated 
with a GRB or an XRF  (Kinugasa et al. 2002; Foley et al. 2003; Valenti et al. 2008).


The broad-line SNe Ic exhibit diversity in terms of the explosion energy, ejecta 
mass and mass of $^{56}$Ni produced during the explosion. The photometric and
spectral features of SN 1998bw are explained with $\sim 10\,\Msun$  
ejected with a kinetic energy $2-5\times$10$^{52}$ ergs, producing 
$0.4 - 0.5 \Msun$  of $^{56}$Ni in the explosion 
(Iwamoto et al. 1998; Nakamura et al. 2001; Maeda et al. 2006, Tanaka et al. 2007). 
 On the other hand, modelling of nebular spectra of SN 2002ap  
indicates  an ejected mass  of $\sim 2.5~\Msun$ with a kinetic
energy $\sim 4$$\times$10$^{51}$ ergs and production of $\sim 0.1\,\Msun$
of $^{56}$Ni (Mazzali et al. 2007), showing a large range in the 
physical parameters of broad-line SNe Ic. 
Among these, SNe with  kinetic energy $E_{\rm K} > 10^{52}$ ergs are 
termed as "hypernovae (HNe)'' (Iwamoto et al. 1998).
The broad-line SNe that are not associated with GRBs are found to
have smaller values of ejecta mass, explosion energy and lower
luminosity as compared to the GRB-associated HNe (Nomoto et al. 2007).

SN 2007ru was discovered by Donati \& Ciabattari on  
November  27.9 and independently by Winslow \& Li with KAIT on November 30.15 
in the spiral galaxy UGC 12381.  
There was no evidence of the supernova in the KAIT image taken on 
November 22.16, down to a limiting magnitude of 18.9 (Donati \& Ciabattari, 
2007). Based on a   spectrum obtained on December 1, SN 2007ru was classified as 
peculiar SN Ic at premaximum phase (Chornock et al. 2007). 
The CaII H \& K and CaII Near Infrared (NIR) triplet absorption troughs were 
found to be weak compared to other SNe Ic.  Further, the OI line at
7774 \AA \ indicated an expansion velocity of 19000 ${\rm km \ sec^{-1}}$, similar to
the expansion velocity seen in the broad-line SN Ic  SN 2006aj (Pian et al. 2006).
A search through the reported discoveries of  GRBs during 
October 15 to November 30, 2007  does not show
any possible association of a GRB with this supernova.

The results of photometric and spectral monitoring of SN 2007ru until around 
210 days after maximum, using the 2m Himalayan Chandra Telescope (HCT) 
of the Indian Astronomical Observatory (IAO), Hanle, India, are presented in
this paper.

\section{Observations and data reduction}
\subsection{Photometry}
SN 2007ru was observed in $UBVRI$ bands during  2007 December 2 (JD 2454437.09)
to 2008 July 03 (JD 2454651.35).  The observations were carried out with the 
Himalayan Faint Object Spectrograph Camera (HFOSC) mounted on the HCT. HFOSC is
equipped with 2K$\times$4K pixels CCD chip. The central 2K$\times$2K region, with a plate
scale of 0.296 arcsec pixel$^{-1}$,  covering a field of 10$^{'}$$\times$10$^{'}$, was 
used for imaging. Gain and readout noise of the CCD camera are 1.22 e$^{-1}$/ADU and 
4.87e$^{-1}$, respectively. Further details on the HCT and HFOSC can be obtained from 
``http://www.iiap.res.in/centers/iao''.

\begin{figure}
\epsscale{1.0}
\plotone{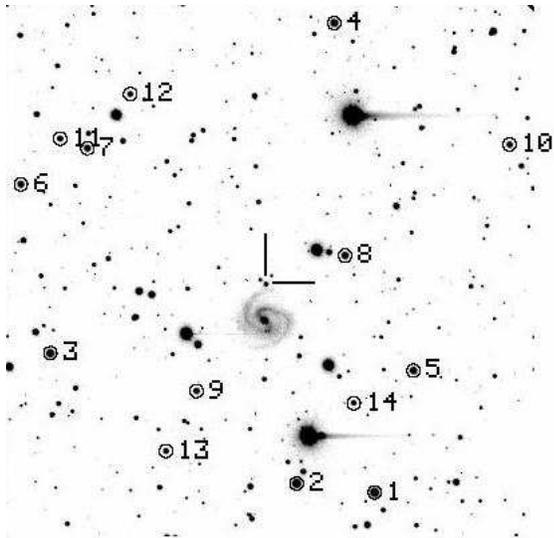}
\caption{Identification chart for SN 2007ru. The stars used as local
standards are marked as numbers 1-14.
\label{fig:field} }
\end{figure}

Photometric standard regions (Landolt 1992)
were observed on 2007 December 25 and December 26 under photometric sky conditions 
to calibrate
a sequence of secondary standards in the supernova field. 
Data reduction has been  done in the  standard manner using various tasks available within 
Image Reduction and Analysis Facility (IRAF). Aperture photometry was performed on the 
photometric standard stars  and
secondary standards, at  an aperture radius
determined using the aperture growth curve. The secondary standard stars
were then calibrated using the average colour terms and the photometric 
zero points determined on the individual night. A sequence of secondary standards
calibrated in this way is marked in Fig. \ref{fig:field} and the $UBVRI$ magnitudes 
of the secondary standards averaged over the two nights are listed in 
Table  \ref{tab:std_star_mag}. 
The magnitudes of the 
supernova and secondary standards were measured using point spread function 
photometry, with a fitting radius equal to the FWHM of the stellar profile. 
Supernova magnitudes were calibrated differentially with respect 
to the local standards. The supernova magnitudes in $U$, $B$, $V$, $R$ and $I$  
bands have been listed in Table \ref{tab:sn_mag}.

\subsection{Spectroscopy}

A series of  spectra were taken during 2007 December 3 to 2008 June 12, 
in the wavelength range (3500\AA\ - 7000\AA) and (5200\AA\ - 9200\AA), with a 
resolution of $\sim$ 7~\AA. The journal of spectroscopic observations is
given in Table  \ref{tab:spec_log}. 
The spectrophotometric standard stars were observed
on the same night to flux calibrate the supernova spectra. 
Spectroscopic data reduction
was carried out using tasks available within IRAF.
The spectra were bias subtracted, flat-fielded  and the one
dimensional spectra were extracted using the optimal extraction method.  The arc lamp
spectra of FeAr and FeNe were used for wavelength calibration. The instrumental 
response correction was effected using the spectroscopic standard spectrum.
The spectra in the two different regions were combined, scaled to a weighted 
mean, to give the  final spectrum on a relative flux scale, which were then 
brought to an absolute flux scale using the $UBVRI$ magnitudes. The supernova 
spectra were corrected for the host galaxy redshift of $z= 0.01546$ 
(from NED) and dereddened by the total reddening $E(B-V) = 0.27$ as
estimated in Section 5.

\section{Optical light curves}
\begin{figure}
\epsscale{1.3}
\plotone{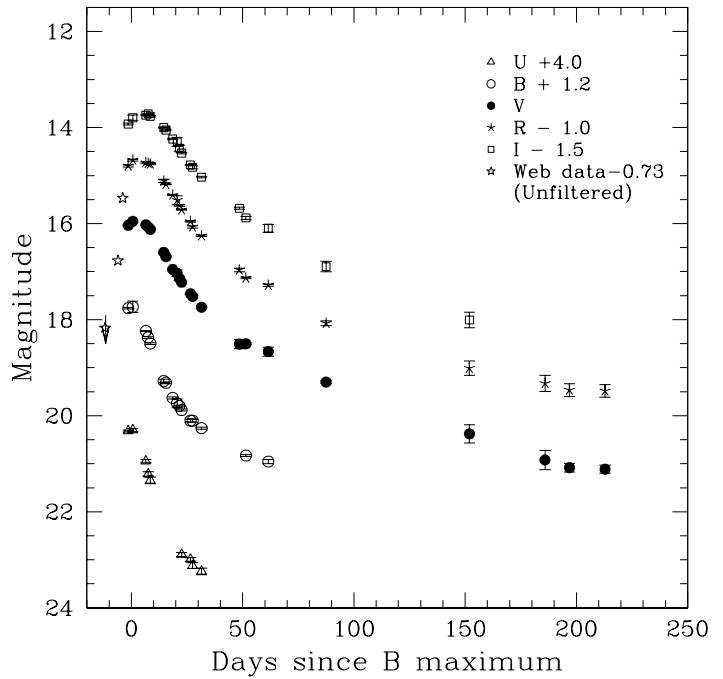}
\caption{The $UBVRI$ \  LCs of SN 2007ru. The LCs have been shifted by the
amount indicated in the legend. The unfiltered magnitudes reported by amatures 
and the pre-discovery limiting magnitudes have been included with $R$ band magnitudes 
in the figure.
\label{fig:light} }
\end{figure}

The $UBVRI$ light curves (LC) of SN 2007ru are shown in 
Fig. \ref{fig:light}. The unfiltered discovery magnitudes and the pre-discovery 
limiting magnitude are also included in the figure. The LCs indicate
that the maximum occurred earlier in the blue, similar to other broad-line SNe Ic. 
The date of explosion can be constrained to $\lsim$ 6 days before discovery (November 25,
JD 2454430$\pm$3), 
based on  the non-detection  on November 22  by KAIT and the subsequent discovery 
on November 27.9.  The rise time to $B$ maximum, which occurred on December 3, is 
5-11 (8$\pm$3) days, indicating SN 2007ru is a fast rising SN Ic 
with a rise time  similar to broad-line SNe  2002ap (Foley  et al. 2003), 
and  2006aj (Modjaz et al. 2006), marginally faster than SN 2003jd (Valenti et al. 2008) but
considerably  faster than the GRB-associated SNe 1998bw and 2003dh (Galama et al. 1998, 
Matheson et al. 2003).   

\begin{figure}
\epsscale{1.3}
\plotone{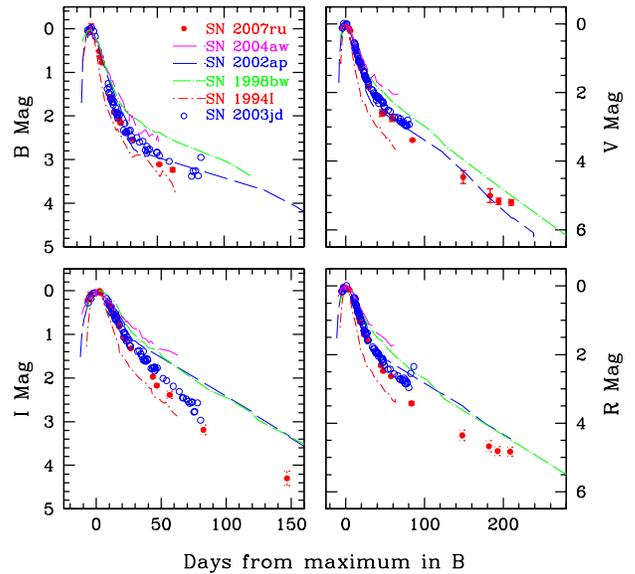}
\caption{Comparison of  LCs of SN 2007ru with other Type Ib/c SNe. The LCs of 
the supernovae in comparison have been shifted arbitrarily to match the date of 
maximum and magnitude at maximum.
\label{fig:fig_comp_lc} }
\end{figure}

The light curves of SN 2007ru are compared  with those of broad-line 
SNe SN 2002ap, SN 2003jd, GRB 980425/SN 1998bw,  normal SNe Ic SN 1994I and 
SN 2004aw in Fig. \ref{fig:fig_comp_lc}. 
A comparison of the decline in brightness 15 days after maximum 
light, $\Delta$m$_{15}$ in different bands shows that SN 2007ru has a  decline similar to 
other broad-line supernovae. In fact, the decline of SN 2007ru is faster than SN 1998bw but 
slower than SNe 2003jd and 2006aj (refer Table  \ref{tab:parameters}). The decline  
rates are estimated to be  $\Delta$m$_{15}$($B$)=1.57,  $\Delta$m$_{15}$($V$)=0.92, 
$\Delta$m$_{15}$($R$)=0.69 
and $\Delta$m$_{15}$($I$)=0.50. The light curves of SN 2007ru decline with 
decline rates of 0.021 mag day$^{-1}$ in $V$, 0.028 mag day$^{-1}$ in $R$ and 
0.030 mag day$^{-1}$ in $I$ bands during days $45-80$. These decline rates
are comparable to the light curve decline rates of the broad-line SN 2003jd and
marginally faster than SN 1998bw. The decline rate in $B$ cannot be estimated
due to a sparse coverage during this period. During the late phases ($>$ 80 
days after explosion), the $V$ and $R$ band light curves of SN 2007ru decline 
with decline rates of 0.0152 and 0.0116 mag day$^{-1}$, respectively, 
slower than both SN 2003jd and SN 1998bw during the corresponding epochs 
(Fig. \ref{fig:fig_comp_lc}). The  decline rate of SN 2007ru during the 
late phases, is faster than  the rate expected due to the radio active decay of  
$^{56}$Co into $^{56}$Fe. This  indicates inefficient trapping of $\gamma$-rays by 
the ejecta, which suggests a low column density.   

The peak absolute magnitudes were estimated using 
the apparent magnitude at maximum in different bands
(see \S 5 for reddening and distance estimate).
 From a comparison of the absolute 
magnitude, SN 2007ru appears to lie at the brighter end of the observed SNe Ic.
 With an absolute $V$
magnitude of $-19.06\pm0.2$, SN 2007ru is fainter than GRB 031203/SN 2003lw 
[$M_{V}$=$-19.75\pm0.5$; 
 Malesani et al. 2004], comparable in brightness 
with SN 1998bw [$M_{V}$=$-19.12\pm0.05$,  Galama et al. 
1998], and brighter than XRF 060218/SN 2006aj [$M_{V}$=$-18.67\pm0.08$;
 Modjaz et al. 2006], broad-line SNe   2002ap [$M_{V}$=$-17.37\pm0.05$;
 Foley et al. 2003, Tomita et al. 2006],  2003jd [$M_{V}$=$-18.9\pm0.3$;
 Valenti et al. 2008] and normal SNe Ic  1994I 
[$M_{V}$=$-17.62\pm0.3$; Richmond et al. 1996, Sauer et al. 2006], 
and  2004aw [$M_{V}$=$-18.02\pm0.3$; Taubenberger et al. 2006].

\section{Spectral evolution}
\begin{figure}
\epsscale{1.3}
\plotone{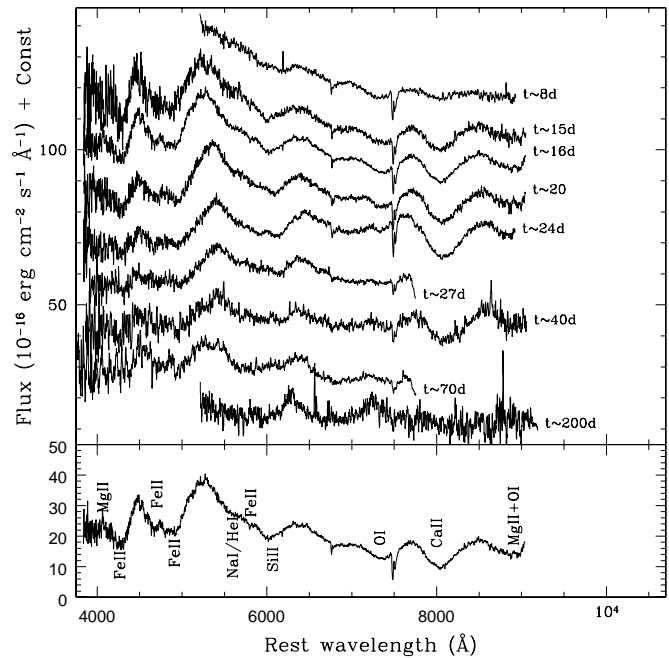}
\caption{Optical spectral evolution of SN 2007ru (top 
panel). The spectra are corrected for the host galaxy redshift. Time in days 
since the day of explosion (2008 November 25- see text) is indicated for each 
spectrum. For clarity, the spectra have been displaced vertically. Main 
spectral features are identified and marked in the spectrum taken $\sim$ 16 
days after explosion (bottom panel).}
\label{fig:spec1}
\end{figure}

The spectral evolution of SN 2007ru is presented in Figure \ref{fig:spec1}. 
The first spectrum, obtained on 2007 December 3, at $t$ $\sim 8$ days, 
where $t$ is days after explosion, 
(the rise time is assumed to be 8 days, see \S 3) 
shows broad absorption features at $\sim 6000$\AA \ due to SiII 
(and a possible contamination as discussed by Valenti et al. 2008), at 
$\sim 7200$\AA \ due to OI and at $\sim 8200$\AA \ due to weak CaII NIR triplet. 
The spectrum at $t$ $\sim 20$ days shows well developed broad absorptions due to MgII 
$\lambda$4481, blends of FeII at $\sim$ 4700\AA, SiII $\lambda$6355, 
OI $\lambda$7774 and CaII NIR triplet. The continuum gets redder by  
$t$ $\sim 20$ days. No significant evolution is seen in the  spectrum taken
at  $t$ $\sim 41$ days. The last spectrum presented here, obtained $\sim 200$
 days after explosion indicates the supernova to be in the nebular phase.  

\begin{figure}
\epsscale{1.3}
\plotone{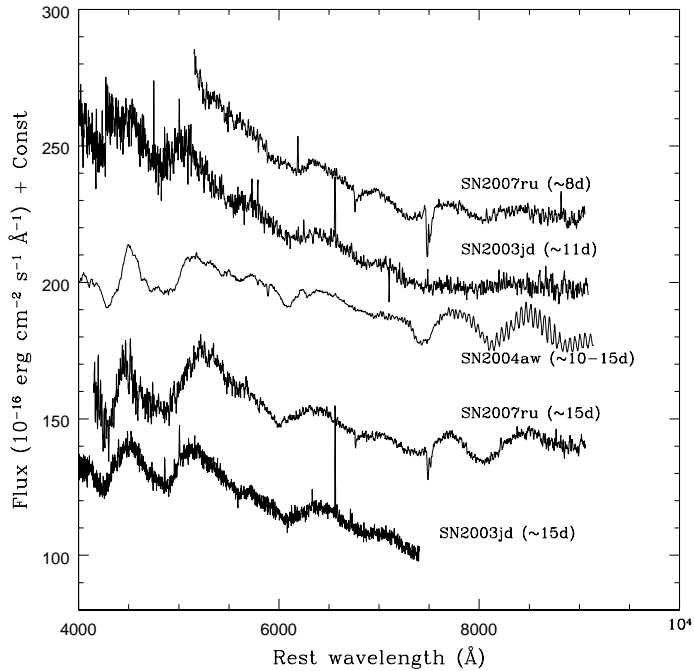}
\caption{Comparison of spectra of SN 2007ru with broad-line type Ic SN 2003jd 
and normal spectrum type Ic SN 2004aw  at different epochs.}
\label{fig:comp_spec1}
\end{figure}

\begin{figure}
\epsscale{1.3}
\plotone{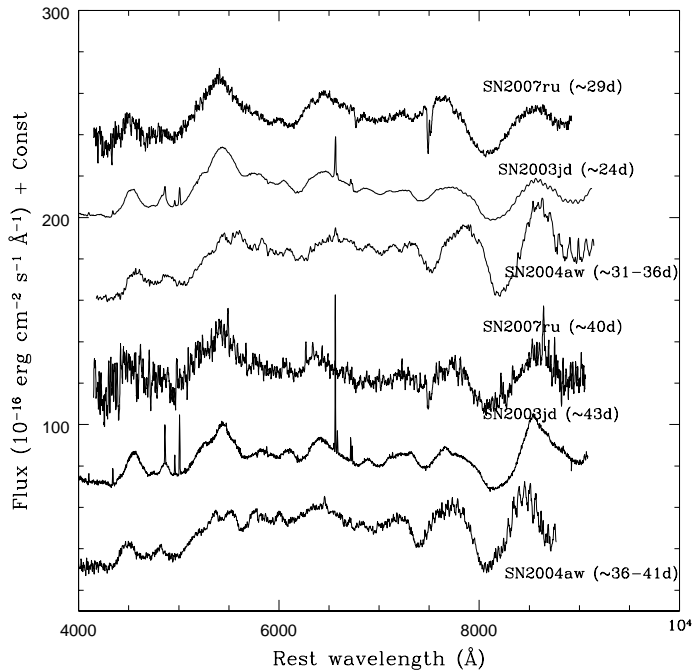}
\caption{Same as figure 5 but at different epochs.}
\label{fig:comp_spec2}
\end{figure}

Comparing the spectrum and its evolution with that of the broad-line SN 2003jd 
and  the 'normal spectrum' Ic supernova SN 2004aw (Fig.\ref {fig:comp_spec1} 
and Fig.\ref {fig:comp_spec2}), 
it is seen that SN 2007ru is very similar to the broad-line SN Ic, indicating high
expansion velocities. The spectral features continue to remain broad and similar to 
those of SN 2003jd even at $t$ $\sim 41$ days.

\subsection{Photospheric velocity}
The high velocity of the supernova ejecta usually results in 
blending of the spectral lines, making direct measurement of
the photospheric velocity difficult. The photospheric velocity of
SN 2007ru is estimated by fitting a Gaussian profile to the 
minimum of the absorption trough of Si II 6355\AA \ 
line, in the redshift corrected spectra. In the first spectrum,
$\sim$8 days after the explosion, the 
absorption feature at 6200\AA \ consists of two components, similar to 
that seen in the premaximum spectra of the broad-line  SN 2003jd 
(Valenti et al. 2008).
Assuming the blue component is due to Si II and the red wing is due to possible 
contamination by other species, the photospheric velocity is estimated to be
20,000 ${\rm km \ sec^{-1}}$. This is consistent with the velocity estimate by Chornock 
et al. (2007) using OI line in the spectrum.

\begin{figure}
\epsscale{1.3}
\plotone{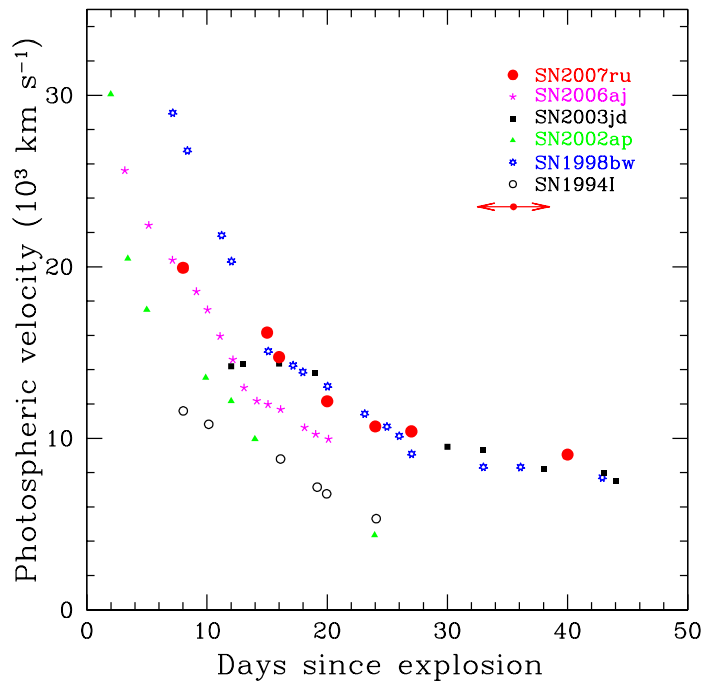}
\figcaption{Evolution of the photospheric velocity of SN 2007ru and other SNe Ic 
inferred from Si II $\lambda$ 6355 line. Uncertainty in estimating days since 
explosion for SN 2007ru due to error in the date of explosion (JD 2454430$\pm$3) is 
indicated by the horizontal arrow.}
\label{fig:phot_vel}
\end{figure}

The photospheric velocity of SN 2007ru,  measured using Si II lines,  and its 
evolution  is compared with other SNe Ic in
Fig \ref{fig:phot_vel}. The photospheric velocity of SN 2007ru at 
$\sim$ 8 days after explosion is lower than GRB 980425/SN 1998bw 
(Patat et al. 2001), but comparable to XRF 060218/SN 2006aj 
(Pian et al. 2006) and broad-line SN 2002ap (Foley et al. 2003). 
However, at the later epochs, the photospheric velocity of SN 2007ru 
is comparable to  those of SN 1998bw and SN 2003jd and  higher than  other
 SNe Ic in comparison. Except for the early phase ($<$ 15 days after explosion)
 the photospheric velocity evolution of SN 2007ru is very similar  to that of  
the broad-line SN 2003jd.   
 
\subsection{Nebular spectrum}
\begin{figure}
\epsscale{1.3}
\plotone{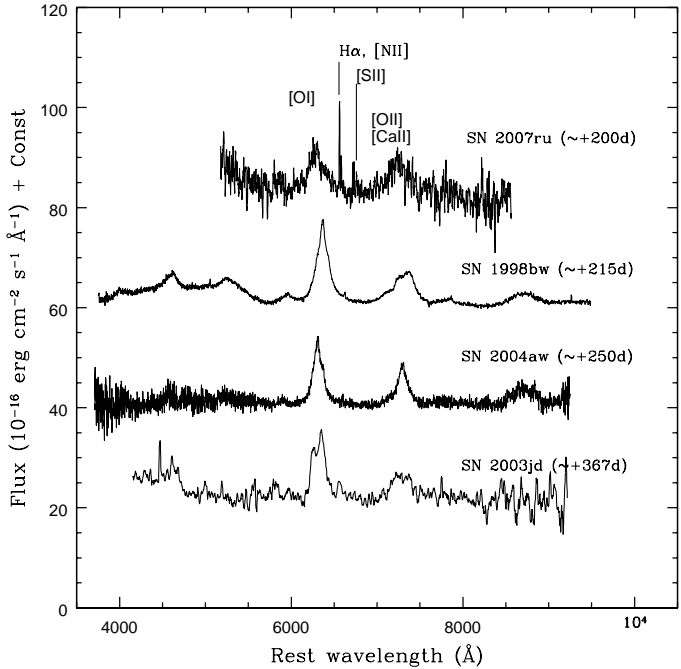}
\caption{Comparison of the nebular spectrum of SN 2007ru with other Type Ic 
supernovae.}
\label{fig:comp_spec_neb}
\end{figure}

The spectrum of SN 2007ru taken $\sim$ 200 days after explosion 
(refer Fig \ref {fig:comp_spec_neb}) is dominated by the  
 forbidden emission lines of [OI] $\lambda$$\lambda$6300, 6364 and
 [CaII] $\lambda$$\lambda$7291, 7323, possibly blended with 
[OII] $\lambda$$\lambda$7320, 7330 (Taubenberger et al. 2006). 
These lines show a broad profile.  
Due to poor signal-to-noise ratio of our spectrum it is difficult to identify 
other spectral lines in the spectrum. However, narrow lines due to H$\alpha$, 
[NII] $\lambda$6583, [SII] $\lambda\lambda$6713, 6731, originating from the 
underlying HII region at the supernova location and  superimposed on the
spectrum of the supernova,  are clearly seen in the spectrum, and are
identified in Fig \ref {fig:comp_spec_neb}.

A comparison of the nebular spectrum of SN 2007ru is made 
with those of other SNe Ic in Fig \ref {fig:comp_spec_neb}. The line profile 
of the [OI] $\lambda$$\lambda$6300, 6364 lines shows a sharp peak, very 
similar to that seen in SN 1998bw, SN 2004aw and SN 2007ru. 
Interestingly, despite the spectral similarity between SN 2007ru and SN 2003jd
at early phase, the [OI] line profile in the nebular spectra is different:
SN 2003jd shows a double peaked structure (see \S 7 for implications).
The profile of [CaII] $\lambda$$\lambda$7291, 7323/[OII] $\lambda$$\lambda$7320, 7330 
line is similar to the [OI] line profile, as seen in SN 2004aw (Taubenberger et al. 2006), 
but different from the profiles seen in SNe 1998bw and 2003jd,  which show a flat topped 
profile. 

The [OI] $\lambda$$\lambda$6300, 6364 and [CaII] $\lambda$$\lambda$7291, 7323 
lines show a blueshift of 2300$\pm$300 ${\rm km \ sec^{-1}}$ and 1200$\pm$200 ${\rm  km \ sec^{-1}}$,
respectively. This could be due to a kinematic offset (Maeda et al. 2007),
optical depth effect (Filippenko et al. 1994),  
or extinction by the dust formed in the SN ejecta (although there is no strong 
indication of the dust formation in the light curve (\S 3))

The velocities derived from full width at half maximum of [OI] and [CaII] lines
are found to be 14000$\pm$2200 ${\rm km \ sec^{-1}}$ and 13500$\pm$1300 ${\rm  km \ sec^{-1}}$, 
respectively. These values are comparable to those seen in SN 1998bw and 
SN 2003jd. The reddening corrected [OI]/[CaII] flux ratio is found to be 
$\sim$ 1.6, which is again comparable to the ratios in SN 1998bw and SN 2003jd.
Thus, 
though the profile of the [OI] and [CaII] lines in the the nebular spectrum 
of SN 2007ru and SN 2003jd differ, other properties like line width and [OI]/[CaII] flux
ratio are similar.

\section{Bolometric light curve}
Direct distance estimates to the host galaxy of SN 2007ru are  not available 
 in the literature. The radial velocity of UGC 12381, corrected for Local 
Group infall onto the Virgo Cluster is 4832 ${\rm km \ sec^{-1}}$  (LEDA). For 
$H_{0}$ = 72 ${\rm km  \ sec^{-1} \ Mpc^{-1}}$, the distance modulus to UGC 12381
is $34.15\pm0.10$, where the error  is  estimated taking into account the 
errors in HI velocity measurement   of the galaxy 
(Paturel et al. 2003)  and the uncertainty in $H_{0}$ .

 The NaID absorption  line is clearly seen in the spectrum with an average
equivalent width of 1.67$\pm$0.37 \AA. Based on the equivalent widths
of  NaID absorption seen in several SNe, Turatto et al. (2003)
find two distinct relations  between  NaID equivalent width and the reddening $E(B-V)$.
Using these relations, the observed NaID equivalent width indicates 
$E(B-V)$ values of 0.85$\pm$0.19 and 0.27$\pm$0.06. The Galactic interstellar 
reddening in the  direction of UGC 12381 is estimated to be 0.26 
(Schlegel, Finkbeiner \& Davis  1998). The NaID absorption seen in the spectra 
of SN 2007ru is clearly from the Milky Way galaxy and no component due to 
the host galaxy is detected. Hence, an $E(B-V)$ value of 0.27 is used for 
extinction correction.


The quasi bolometric LC of SN 2007ru is estimated using the 
$UBVRI$ magnitudes corrected for reddening with $E(B-V)$=0.27 and the Cardelli et al. (1989) 
extinction law. The  magnitudes were  converted to the monochromatic 
flux, using zero points from  Bessell et al. (1998). The fluxes were then spline 
interpolated and integrated from 
3100\AA\ to 1.06$\mu$m. Since $U$ band observations are not available beyond 
35 days since $B$ maximum, the bolometric LC is estimated by 
integrating the $BVRI$ fluxes only. The contribution of $U$ band to the
bolometric flux at phases $\sim  35$ days is estimated to be $\lsim 10\%$. 
In the later phases when only $V$, $R$, $I$ or $V$, $R$ magnitudes are 
available,
the bolometric magnitudes are derived by applying a bolometric correction
to the available magnitudes. The bolometric corrections were estimated based 
on the last four points for which  $B$, $V$, $R$ and $I$ measurements are 
available.

\begin{figure}
\epsscale{1.0}
\plotone{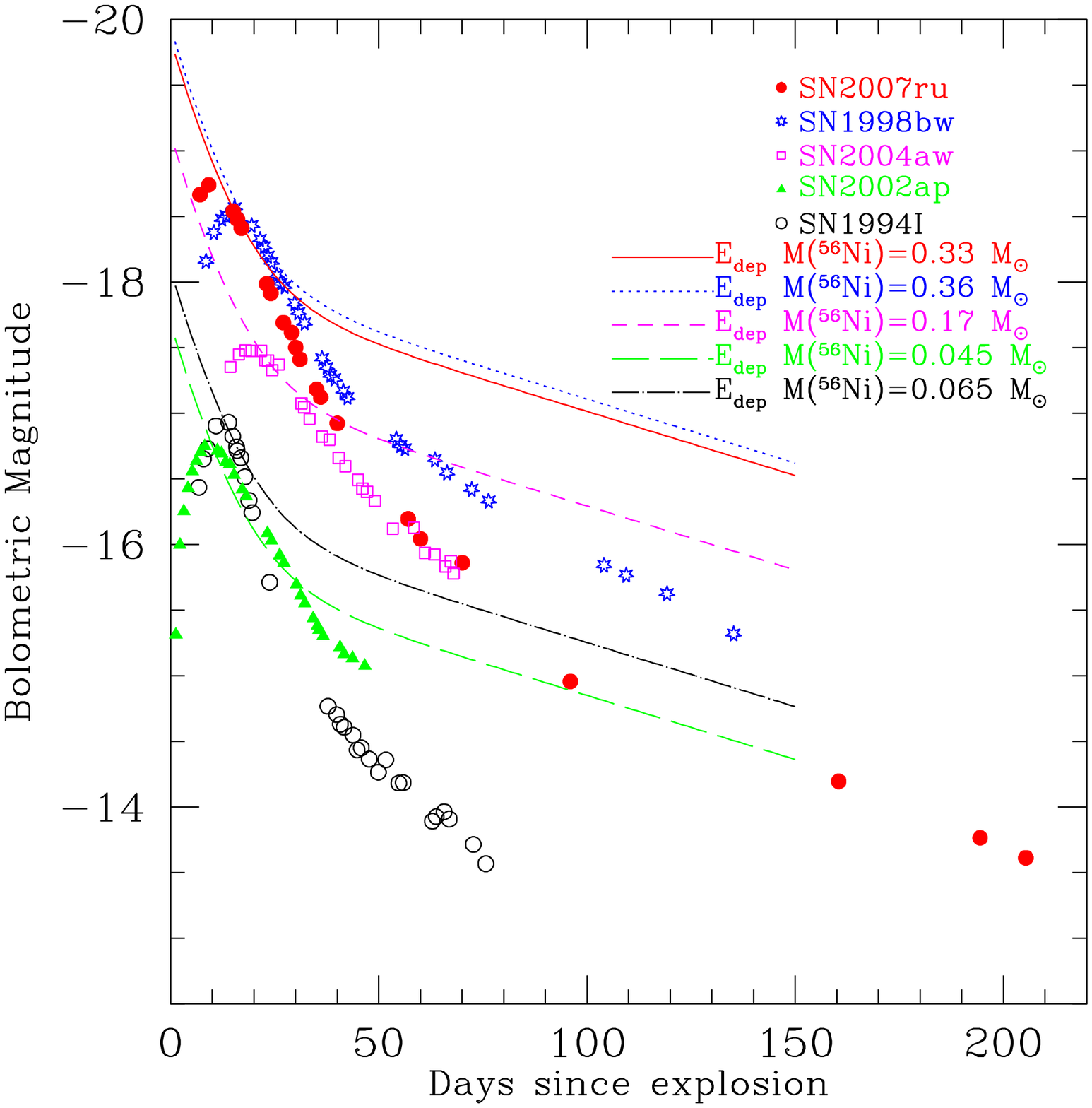}
\figcaption{Quasi bolometric  light curves of SN 2007ru, SN 1998bw, SN2002ap, 
SN1994I and SN 2004aw, estimated as explained in text. The  lines 
represent the  rate of energy production via 
 \Nifs --  \Cofs \ chain for different values of M(\Nifs).
\label{fig:light_bol}}
\end{figure}

 The quasi bolometric LC  is shown in Fig \ref{fig:light_bol}.
Adding a conservative uncertainty of $\pm0.2$, the bolometric
magnitude at maximum is estimated as $-18.78\pm0.2$. The quasi bolometric
LCs, estimated in a similar manner, for SN 1998bw, SN 2002ap, 
SN 2004aw and SN 1994I,  are also plotted in Fig. \ref{fig:light_bol}. 
The quasi bolometric LC of SN 2007ru is brighter than the other well studied  
non-GRB  broad-line SN 2002ap and normal SNe Ic,
and comparable to SN 1998bw. The decline in the 
bolometric LC of SN 2007ru is considerably faster than SN 1998bw and comparable
to SN 2002ap.

 Using  Arnett's rule (Arnett 1982), 
the mass of \ \Nifs \  required to power the quasi bolometric LC of SN 2007ru
is estimated to be $0.33 \Msun$, whereas it is $0.36 \Msun$ for SN 1998bw 
(Fig. \ref{fig:light_bol}). It is to be noted here that the contribution due to 
NIR bands is not included in the bolometric LC. The
NIR contribution to bolometric flux  for broad-line SNe  2002ap and  1998bw is 
$\sim 30\%$ (Tomita et al. 2006, Valenti et al. 2008), whereas for SN 1994I it 
is only $\sim$ 10$\%$, while for SN 2004aw the NIR  contribution
increases from $\sim$ 31$\%$ to $\sim$ 45$\%$ between +10
and +30 days  (Taubenberger et al. 2006). Assuming an  NIR 
contribution to the bolometric flux similar to SNe 2002ap and 1998bw ($\sim$ 30$\%$),  
 the mass of \ \Nifs\ for SN 2007ru is estimated to be  $\sim$ 0.4 $\Msun$. 
The total rate of energy production via \ \Nifs - 
\Cofs \ chain  estimated using the analytical formula by Nadyozhin (1994), for
different values of mass of \ \Nifs \  synthesized during the explosion and are 
plotted with the quasi bolometric LC (thin lines) in Fig \ref{fig:light_bol}.
The plots indicate a good match of the energy production rate for $0.33 \Msun$ 
of \ \Nifs \  with the initial decline of the quasi-bolometric light curve of SN 2007ru, 
in agreement with the estimate based on Arnett's rule.

\section{Properties of the host galaxy of SN 2007ru}

\subsection{The supernova region}
An attempt is made to estimate the metallicity of the region where the 
supernova exploded, based on the observed [NII]/H$\alpha$ flux ratio, from the
underlying HII region, superimposed  in the 
nebular spectrum of the supernova. Following Pettini \& Pagel (2004), the 
N2 index (log[NII]/H$\alpha$) is estimated to be $-0.36$. Using this, an 
oxygen abundance of $12 + \log (O/H) = 8.78$ is derived. Another way
of deriving oxygen abundance is using the $\log ([NII]/H\alpha)$ diagnostic 
diagram (Kewley \& Dopita, 2002), which requires an estimate of the ionization 
parameter $q$ or $U$ ($U =q/c$; $c$ is the speed of light) also. The ionization 
parameter {\it U} can be estimated from the [SII]/[SIII] ratio following 
Diaz et al. (1991). The observed flux of [SII] lines $\lambda$$\lambda$6717, 
6731 and [SIII] line $\lambda$9069, seen in the nebular spectrum can be used 
to estimate the ionization parameter, however, our spectrum does not cover the  
[SIII] $\lambda$9532 region. In the extragalactic HII regions  ratio 
([SIII]$\lambda$9532/[SIII]$\lambda$9069) is found to vary in the range 
1.58 to 3.77  with an observed mean of 2.66$\pm$0.46,  against the theoretical 
value of 2.44 (Diaz et al. 1985, Vilchez et al. 1988, Kennicutt \& Garnett 1996)   
This indicates that the ionization parameter $q$ can vary in the range 
$\sim 10^{6}$ to 3$\times10^{6}$.  Using the 
log([NII]/H$\alpha$) diagnostic diagram for the above estimated range of the 
ionization parameter, the oxygen abundance 12 + {log}(O/H)  is found to lie close to 
8.8, which is in good agreement with the   independent estimate of 12 + {log}(O/H) = 8.78.
This indicates the oxygen abundance in region where the supernova SN 2007ru occurred is 
close to solar. 

In a recent study Modjaz et al. (2008b) have concluded  that the broad-line 
supernovae associated with GRBs are generally found in  metal poor  environments as compared 
to the broad-line supernovae without GRBs. They have shown that,  in their 
sample,  the 
oxygen abundance 12 + log(O/H)$_{KD2}$ = 8.5 can be treated as the boundary 
between galaxies that have GRBs associated supernovae and
those without GRBs.  The estimated oxygen abundance of $\sim$ 8.8  at the 
location of SN 2007ru fits well in the range expected for a broad-line 
supernova without GRB.    

\subsection{The nuclear region}

\begin{figure}
\epsscale{1.0}
\plotone{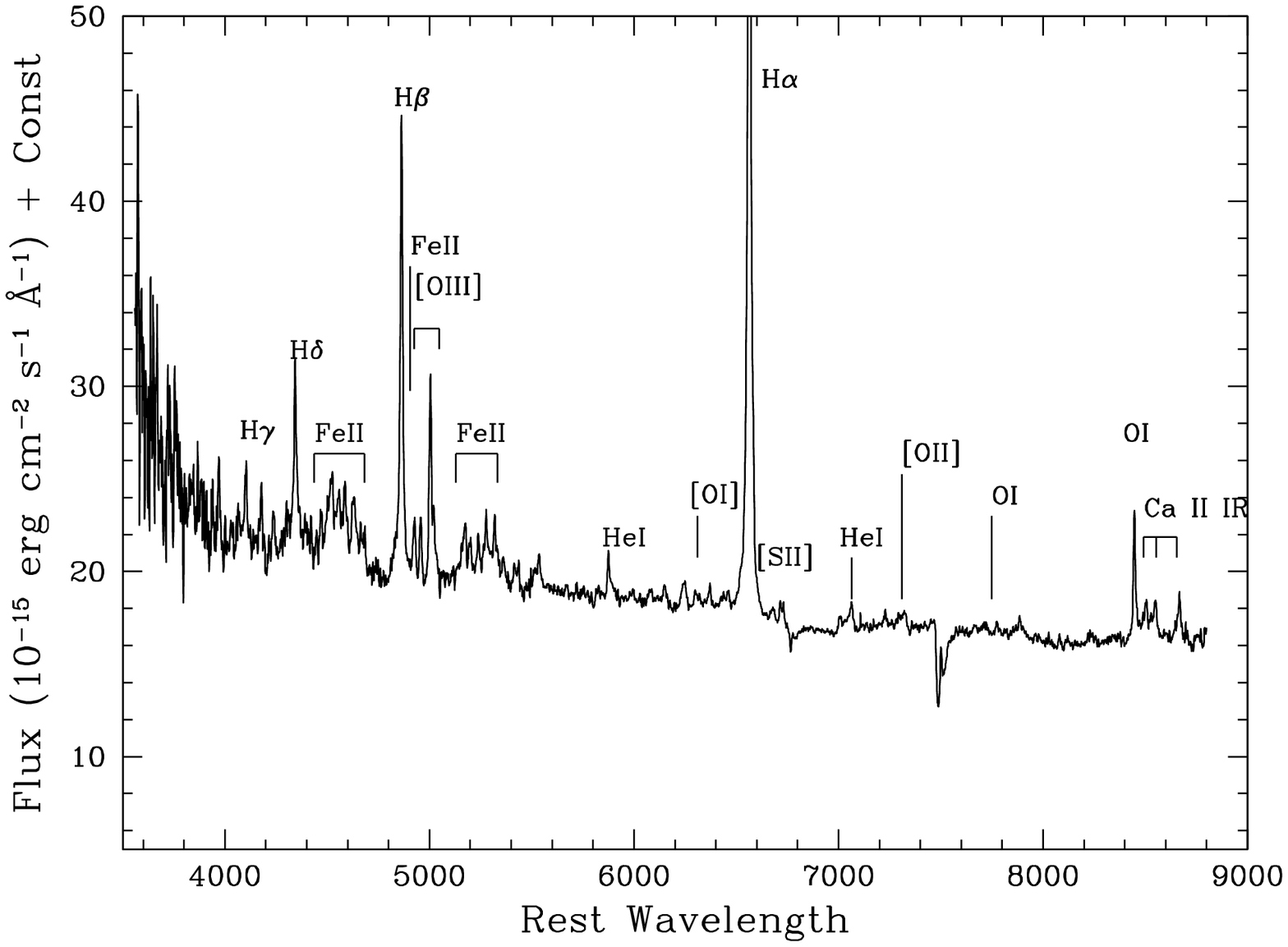}
\figcaption{Nuclear spectrum of the host galaxy UGC 12381 of SN 2007ru.
\label{fig:nuc_spec}}
\end{figure}

The nuclear spectrum of the host galaxy of SN 2007ru (obtained on 2008 October 
29) is shown in Fig. \ref{fig:nuc_spec}. The nuclear spectrum shows 
strong hydrogen lines of the Balmer series, permitted as well as forbidden 
lines of oxygen, lines due to helium and the calcium NIR triplet  
(refer Fig. \ref{fig:nuc_spec}).  The FWHM velocities of the lines indicate 
velocities of the order of 500-700 ${\rm km \ sec^{-1}}$. The hydrogen Balmer lines show 
broad wings, with a noticable asymmetry in the blue wing. Such broad wings are
not seen in the forbidden lines. Another interesting feature of the 
nuclear spectrum of the host galaxy is the presence of numerous FeII lines 
seen at wavelengths 4400--4600\AA, 4924\AA, 5018\AA \ and 5100--5400\AA, 
similar to the spectra of  Active Galactic Nuclei (Veron-Cetty et al. 2004, 
Veron-Cetty et al. 2006).

Using the diagnostic diagram (Ho, Filippenko \& 
Sargent 1997) based on the line ratio $\log([OIII]\lambda 5007/H\beta)$ versus 
$\log([OI]\lambda 6300/H\alpha)$ \ and \ $\log([OIII]\lambda5007/H\beta)$ 
versus  \ \ $\log([SII]\lambda\lambda6717,6731/H\alpha)$  the nucleus of UGC 12381 
can be classified as belonging to the HII region class, with very weak [SII] and [NII] lines.


 It thus appears that the host galaxy of SN 2007ru probably hosts a mild 
AGN with a nuclear HII region, which needs further detailed study.

\section{Discussion \& Conclusions}

 The optical spectra of SN 2007ru presented here show broad spectral features
similar to that seen in GRB-associated SN 1998bw, XRF-associated SN 2006aj and the
broad-line SN 2003jd. The expansion velocity of the ejecta of SN 2007ru is higher 
than that of  the normal SN Ic SN 1994I and SN 2006aj, but comparable to that of 
SN 1998bw and SN 2003jd.

The maximum luminosity of SN 2007ru is comparable to SN 1998bw.
The light curve reaches a peak in only $\sim 8\pm 3$ days, which is remarkably 
faster than in SN 1998bw ($\sim$ 20 days).
The mass of \Nifs\ ejected in SN 2007ru is estimated as $\sim 0.4 \Msun$,
which is similar to that estimated for SNe 1998bw and 2003jd, 
slightly larger than that for SN 2004aw, and 
much larger than that for the broad-line SNe 2002ap and 2006aj 
and the normal SN Ic 1994I (Table  \ref{tab:parameters}).

From the rise time of the light curve ($\tau$) and the expansion velocity ($v$),
we  estimate  mass of SN ejecta ($M_{\rm ej}$) and the kinetic energy of the ejecta 
($\KE$). If the optical opacity is assumed to be constant, the timescale of 
the light curve is expressed as $\tau \propto \Mej^{3/4} \KE^{-1/4}$ 
(Arnett 1982). The expansion velocity is given by 
$v \propto \Mej^{-1/2} \KE^{1/2}$.
The rise time of SN 2007ru ($8 \pm 3$ days) is comparable to that of the
well studied SN 1994I, while the expansion velocity ($v = 20,000$ ${\rm km \ sec^{-1}}$ at 
maximum) is about twice. Assuming $\Mej=1.0 \Msun$ and $\KE=1.0 \times 10^{51}$
ergs for SN 1994I (see Table \ref{tab:parameters}), and the observed expansion velocity of
SN 2007ru, we estimate $\Mej =1.3^{+1.1}_{-0.8} \Msun$ and 
$\KE = 5^{+4.7}_{-3.0} \times 10^{51}$ ergs for SN 2007ru. A similar analysis 
for SN 1998bw, assuming the rise time and the velocity of SN 1998bw to be 
twice that of SN 1994I, leads to  $\Mej \approx 8 \Msun$ and 
$\KE \approx 30 \times 10^{51}$ergs,
qualitatively consistent with the results of detailed modelling
(Iwamoto et al. 1998; Nakamura et al. 2001)
\footnote{Given the very high
$M_{\rm Ni}/M_{\rm ej}$ ratio ($\sim 0.3$), the higher end of
the ejecta mass may be preferred ($\Mej \sim 2.4 \Msun$,
and $\KE \sim 9.7 \times 10^{51}$ erg).
Since the explosion with a larger kinetic energy can produce a larger
amount of \Nifs, the large \Nifs\ mass in SN 2007ru may also support this.}.
If we take SN 2003jd as a reference (Valenti et al. 2008),
$\Mej =1.7^{+1.5}_{-1.0} \Msun$ and
$\KE = 8.6^{+7.2}_{-5.2} \times 10^{51}$ ergs are derived for SN 2007ru.
It should  however be noted, that a
detailed modelling is required  to derive  accurate values of $\Mej$ and $\KE$
\footnote{An attempt was made to estimate the mass of the ejecta $\Mej$ and 
kinetic energy $\KE$ using the width of the light curve. The width is defined 
as the time from peak of the bolometric light curve to the time when the luminosity 
is equal to the 1/{\it e} times the peak luminosity, which is equivalent to  
a decline of 1.1 mag from peak. The derived width is $\sim$~18 days for SN 2007ru.
With this value, we have estimated $\Mej$  and $\KE$ in the same manner
and derived larger values of $\Mej$  and $\KE$ ($M_{\rm ej} \sim 4.5 \Msun$ 
and $E_{\rm K} \sim 20 \times 10^{51}$ erg). Our conclusion of large E/M is thus not affected.}.

SN 2007ru has a large kinetic energy while the ejecta mass is close to that of 
normal SNe Ic. Studies of SNe Ic (\eg Nomoto et al. 2007) have shown a trend, 
although weak, wherein SNe having massive ejecta tend to have a larger kinetic 
energy and eject more \Nifs, connecting normal SNe to GRB-associated SNe.  
In contrast to this trend, SN 2007ru which resides at the higher energy end has
a lower mass ejecta, leading to a higher $E/M$.
SN 2007ru thus adds to the diversity of SNe Ic.

The spectroscopic properties of SN 2007ru at early phases are
most similar to SN 2003jd.
Also if we take the higher end of $\Mej$, the ejecta properties
of SN 2007ru are also close to those of SN 2003jd
(although the estimated $E/M$ is higher for SN 2007ru).
However, the [OI] line profiles in the nebular spectra
are dissimilar.  SN 2007ru shows a single peaked profile while
SN 2003jd shows a double-peaked profile.
In aspherical explosions, we would expect a single-peaked [OI] profile
for the polar-viewed case, and a double-peaked profile for the side-viewed case
(Mazzali et al. 2005; Maeda et al. 2008; Modjaz et al. 2008a).
Also, the polar-viewed aspherical explosion tends to show a brighter
peak (Maeda et al. 2006) and faster velocity (Tanaka et al. 2007).
This  matches with the properties of SN 2007ru.
Thus, we suggest that SN 2007ru could be an aspherical explosion viewed from
the polar direction.
Detailed multidimensional modelling is required to answer if the high $E/M$
derived for SN 2007ru results from the effect of asphericity.

The nebular spectrum of SN 2007ru shows narrow emission lines due to H$\alpha$,
[NII], [SII] and [SIII], arising from the underlying/neighbouring host
galaxy HII region. The flux ratios indicate an oxygen abundance
of $\sim$ 8.8 in the region of the supernova. The nearly solar oxygen abundance 
at the location of the supernova matches
well with earlier abundance studies for supernova host galaxies. 

The nuclear spectrum of the host galaxy of SN 2007ru shows broad hydrogen 
Balmer lines, with an asymmetric blue wing. Emission lines due to FeII are also 
fairly prominent. Low ionization emission lines are also present. It appears
that the galaxy hosts a mild AGN with nuclear HII region.

\acknowledgements
We are thankful to the anonymous referee for valuable comments and T.P. Prabhu
for useful discussions. This work has
been carried out under the INSA (Indian National Science Academy) -
JSPS (Japan Society for Promotion of Science) exchange programme.
This research was supported in part by the National Science Foundation
under Grant No. PHY05-51164,   World Premier International 
Research Center Initiative, MEXT, Japan, and by the Grant-in-Aid 
for Scientific Research of the JSPS (18104003, 18540231, 20540226) 
and MEXT (19047004, 20040004). M.T. is  supported through the JSPS Research 
Fellowships for Young Scientists. We thank all the observers of the
2-m HCT (operated by the Indian Institute of Astrophysics), who kindly
provided part of their observing time for the supernova observations.
This work has made use of the NASA Astrophysics Data System and the NASA/IPAC
Extragalactic Database (NED) which is operated by Jet Propulsion Laboratory,
California Institute of Technology, under contract with the National
Aeronautics and Space Administration.

\begin{deluxetable}{lccccc}
\tablewidth{0pt}
\tablecaption{Magnitudes for the sequence of secondary standard stars in
the field of SN 2007ru.\tablenotemark{*}}
\tablehead{
ID & U  & B & V &  R & I
}
\startdata
1 &   $14.433\pm0.015  $& $14.396\pm0.006 $&  $13.877\pm0.005  $&    $13.531\pm0.005 $&  $13.116\pm0.006  $ \\
2 &   $14.918\pm0.017  $& $14.790\pm0.013 $&  $14.073\pm0.018  $&    $13.618\pm0.016 $&  $13.111\pm0.023  $ \\
3 &   $15.111\pm0.020  $& $15.020\pm0.016 $&  $14.343\pm0.020  $&    $13.953\pm0.017 $&  $13.488\pm0.024  $ \\
4 &   $15.870\pm0.030  $& $15.554\pm0.010 $&  $14.714\pm0.021  $&    $14.232\pm0.012 $&  $13.689\pm0.019  $ \\
5 &   $16.105\pm0.030  $& $15.670\pm0.021 $&  $14.796\pm0.024  $&    $14.317\pm0.002 $&  $13.783\pm0.004  $ \\
6 &   $15.390\pm0.025  $& $15.339\pm0.014 $&  $14.780\pm0.022  $&    $14.406\pm0.021 $&  $13.995\pm0.021  $ \\
7 &   $15.773\pm0.032  $& $15.616\pm0.019 $&  $14.869\pm0.024  $&    $14.444\pm0.024 $&  $13.933\pm0.030  $ \\
8 &   $16.138\pm0.030  $& $15.900\pm0.012 $&  $15.046\pm0.021  $&    $14.550\pm0.023 $&  $13.985\pm0.030  $ \\
9 &   $16.792\pm0.036  $& $16.136\pm0.018 $&  $15.139\pm0.022  $&    $14.553\pm0.014 $&  $13.951\pm0.025  $ \\
10&   $17.312\pm0.045  $& $16.619\pm0.017 $&  $15.631\pm0.017  $&    $15.066\pm0.014 $&  $14.471\pm0.014  $ \\
11&   $17.321\pm0.040  $& $16.730\pm0.014 $&  $15.739\pm0.025  $&    $15.153\pm0.014 $&  $14.555\pm0.021  $ \\
12&   $17.970\pm0.052  $& $16.954\pm0.019 $&  $15.766\pm0.025  $&    $15.083\pm0.024 $&  $14.362\pm0.031  $ \\
13&   $16.552\pm0.045  $& $16.505\pm0.020 $&  $15.796\pm0.030  $&    $15.368\pm0.026 $&  $14.864\pm0.030  $ \\
14&   $17.518\pm0.050  $& $16.997\pm0.005 $&  $16.045\pm0.018  $&    $15.513\pm0.018 $&  $14.937\pm0.027  $ \\

\enddata
\tablenotetext{*}{The stars are identified in Figure. \ref{fig:field}}
\label{tab:std_star_mag}
\end{deluxetable}

\begin{deluxetable}{lccccccc}
\tablewidth{0pt}
\tablecaption{Photometric observations of SN 2007ru}
\tablehead{
 Date & J.D. & Phase\tablenotemark{*} & U & B & V & R & I\\
     & 2454000+ & (days)    &   &   &   &   &}
\startdata
02/12/2007&   437.086& -2& $16.318\pm0.021$ & $ 16.562\pm0.016$ &  $16.036\pm0.011$ & $ 15.798\pm0.022$ & $ 15.426\pm0.020$\\
04/12/2007&   439.155&  0& $16.299\pm0.029$ & $ 16.533\pm0.120$ &  $15.951\pm0.015$ & $ 15.678\pm0.019$ & $ 15.296\pm0.056$\\
10/12/2007&   445.029&  6& $16.943\pm0.034$ & $ 17.036\pm0.008$ &  $16.025\pm0.012$ & $ 15.726\pm0.012$ & $ 15.244\pm0.014$\\
11/12/2007&   446.047&  7& $17.211\pm0.045$ & $ 17.172\pm0.019$ &  $16.070\pm0.011$ & $ 15.746\pm0.008$ & $ 15.215\pm0.011$\\
12/12/2007&   447.027&  8& $17.346\pm0.067$ & $ 17.296\pm0.019$ &  $16.121\pm0.011$ & $ 15.759\pm0.014$ & $ 15.265\pm0.016$\\
18/12/2007&   453.046&  14&                 & $ 18.078\pm0.039$ &  $16.598\pm0.014$ & $ 16.111\pm0.035$ & $ 15.501\pm0.010$\\
19/12/2007&   454.056&  15&                 & $ 18.114\pm0.019$ &  $16.687\pm0.012$ & $ 16.184\pm0.007$ & $ 15.553\pm0.017$\\
22/12/2007&   457.067&  18&                 & $ 18.430\pm0.024$ &  $16.954\pm0.011$ & $ 16.404\pm0.011$ & $ 15.738\pm0.013$\\
24/12/2007&   459.097&  20&                 & $ 18.540\pm0.090$ &  $17.034\pm0.070$ & $ 16.512\pm0.095$ & $ 15.799\pm0.095$\\
25/12/2007&   460.069&  21&                 & $ 18.596\pm0.017$ &  $17.140\pm0.035$ & $ 16.621\pm0.016$ & $ 15.929\pm0.060$\\
26/12/2007&   461.079&  22& $18.889\pm0.040$ & $ 18.681\pm0.021$ &  $17.221\pm0.024$ & $ 16.711\pm0.010$ & $ 16.031\pm0.016$\\
30/12/2007&   465.061&  26& $18.990\pm0.038$ & $ 18.902\pm0.032$ &  $17.460\pm0.008$ & $ 16.956\pm0.013$ & $ 16.278\pm0.008$\\
31/12/2007&   466.029&  27& $19.118\pm0.060$ & $ 18.902\pm0.036$ &  $17.517\pm0.014$ & $ 17.069\pm0.024$ & $ 16.329\pm0.031$\\
04/01/2008&   470.045&  31& $19.241\pm0.069$ & $ 19.057\pm0.021$ &  $17.741\pm0.024$ & $ 17.252\pm0.015$ & $ 16.532\pm0.016$\\
21/01/2008&   487.041&  48&                  &                   &  $18.509\pm0.096$ & $ 17.964\pm0.037$ & $ 17.181\pm0.022$\\
24/01/2008&   490.063&  51&                  & $ 19.629\pm0.021$ &  $18.501\pm0.019$ & $ 18.130\pm0.014$ & $ 17.382\pm0.031$\\
03/02/2008&   500.055&  61&                  & $ 19.754\pm0.044$ &  $18.665\pm0.092$4& $ 18.282\pm0.023$ & $ 17.597\pm0.082$\\
29/02/2008&   526.068&  87&                  &                   &  $19.300\pm0.026$ & $ 19.075\pm0.050$ & $ 18.395\pm0.109$\\
03/05/2008&   590.431& 151&                  &                   &  $20.377\pm0.19 $ & $ 20.007\pm0.15 $ & $ 19.506\pm0.164$\\
06/06/2008&   624.400& 185&                 &                   &  $20.921\pm0.20 $ & $ 20.325\pm0.169$ & \\
17/06/2008&   635.360& 196&                 &                   &  $21.083\pm0.088$ & $ 20.466\pm0.133$ & \\
03/07/2008&   651.348& 212&                 &                   &  $21.115\pm0.084$ & $ 20.483\pm0.128$ & \\ 
\enddata  		     
\tablenotetext{\rlap{*}}{{ \,Observed phase with respect to the epoch of
maximum in $B$ band (JD 2454438.8).}}
\label{tab:sn_mag}
\end{deluxetable}

\begin{deluxetable}{lcrc}
\tablewidth{0pt}
\tablecaption{Log of spectroscopic observations of SN 2007\lowercase{ru}}
\tablehead{
Date & J.D. & \multicolumn{1}{c}{Phase\rlap{*}} & Range \\
     & 2454000+ & \multicolumn{1}{c}{(days)} & \AA\  \\
}
\startdata
03/12/07 & 438.21 & $ +8  $   &  5200-9100 \\          
10/12/07 & 445.06 & $ +15 $   & 3500-7000; 5200-9100\\ 
11/12/07 & 446.07 & $ +16 $   & 3500-7000; 5200-9100\\ 
15/12/07 & 450.14 & $ +20 $   & 3500-7000; 5200-9100\\ 
19/12/07 & 454.13 & $ +24 $   & 3500-7000; 5200-9100\\ 
22/12/07 & 457.09 & $ +27 $   & 3500-7000           \\ 
04/01/08 & 470.09 & $ +40 $   & 3500-7000; 5200-9100\\ 
03/02/08 & 500.08 & $ +70 $   & 3500-7000           \\  
12/06/08 & 630.38 & $ +200 $   & 5200-9100 \\          

\enddata
\tablenotetext{*}{Relative to the epoch of date of explosion  JD = 2454430}
\label{tab:spec_log}
\end{deluxetable}

\begin{deluxetable}{lllllllllll}
\tablewidth{0pt}
\tablecaption{Comparison of parameters of SNe~Ic }
\tablehead{          
& M$_V$ &$\Delta$m$_{15}$(V)& $\gamma_{\rm V}$&$\gamma_{\rm R}$&$\gamma_{\rm I}$&$E_{\rm K} / 10^{51}$ergs & $M_{\rm ej}/M_\odot$ & $M_{\rm N\!i}/M_\odot$ & $E_{\rm K}/M_{\rm ej}$ & References 
}

\startdata

SN 1994I       & $-17.62$ &1.65 &0.029 &0.028 &0.026  & 1                      & 0.9--1           & 0.07(0.07)     & $\sim 1$         & 1, 2, 3 \\
SN 2004aw      & $-18.02$ &0.62 &0.014 &0.017 &0.015  & 3.5--9.0               & 3.5--8.0         & 0.25--0.35(0.2)  & $\sim 1$      &  4 \\
SN 2003jd      & $-18.9$  &1.44 &0.022 &0.022 &0.029  & 7$^{+3}_{-2.0}$        & 3.0$\pm$1.0              & 0.36        & $\sim 2.3$  & 5 \\
SN 2002ap      & $-17.35$ &0.87 & & &  & 4                      & 2.5                & 0.1(0.06)  & $\sim1.6$      & 6, 7, 8 \\
SN 2007ru      & $-19.06$ &0.92 &0.021 &0.028 &0.030  & 5$^{+4.7}_{-3.0}$      & 1.3$^{+1.1}_{-0.8}$  & 0.4     & $\sim 3.8$         & This work\\
SN 1998bw      & $-19.13$ &0.75 &0.020 &0.022 &0.022  & 30                     & 10               & 0.40(0.5)         & $\sim 3$     & 9, 10, 11\\
SN 2006aj       & $-18.7$ &1.14 & & &  & 2                      & 2                & 0.21        & $\sim 1$     & 12, 13\\
\enddata

\tablenotetext{*}{$M_{\rm N\!i}$ given in  parenthesis are the values estimated using the Arnett's rule (including NIR contribution in the bolometric
 light curves, see text), $\gamma$ represents the magnitude decline rates (mag/day) between  45 - 80 days}
\tablerefs{(1) Richmond et al. 1996; (2) Nomoto et al. 1994; (3) Sauer et al. 2006; (4) Taubenberger et al. 2006; (5) Valenti et al. 2008;
                        (6) Mazzali et al. 2007;  (7) Foley et al. 2003;  (8) Tomita et al. 2006; (9) Galama et al. 1998; (10) Iwamoto et al. 1998;
                        (11) Nakamura et al. 2001; (12) Modjaz et al. 2006; (13) Mazzali et al. 2006 }
\label{tab:parameters}
\end{deluxetable}

\end{document}